\documentclass[preprint2]{aastex}
\usepackage{graphicx}

\newcommand{\chandra}{{\it Chandra} }
\newcommand{\xmm}{{\it XMM-Newton} }

\slugcomment{Not to appear in Nonlearned J., 45.}

\shorttitle{AGN activity in fossil galaxy cluster RX
J1416.4+2315}
\shortauthors{Miraghaei et al.}

\begin{document}


\title{AGN activity and IGM heating in fossil cluster RX
J1416.4+2315}

\author{H. Miraghaei\altaffilmark{1,2},
H. G. Khosroshahi\altaffilmark{1}, 
C. Sengupta\altaffilmark{3},
S. Raychaudhury\altaffilmark{4,5},
N. N. Jetha\altaffilmark{6},
S. Abbassi\altaffilmark{7,1}}

\email{halime@ipm.ir}

\altaffiltext{1}{School of Astronomy, Institute for Research in
Fundamental Sciences, PO Box 19395-5531, Tehran, Iran}
\altaffiltext{2}{Department of Physics, Sharif University of
Technology PO Box 11365-9161, Tehran, Iran}
\altaffiltext{3}{Korea Astronomy and Space Science Institute,
776, Daedeokdae-ro, Yuseong-gu, Daejeon, Republic of Korea
305-348}
\altaffiltext{4}{Department of Physics, Presidency University, 86/1 College Street, 700 073 Kolkata, India}
\altaffiltext{5}{School of Physics and Astronomy, University of
Birmingham, Edgbaston, Birmingham B15 2TT, UK}
\altaffiltext{6}{Center for Space Plasma and Aeronomic Research
(CSPAR), University of Alabama in Huntsville, Huntsville, AL
35805, USA}
\altaffiltext{7}{Department of Physics, School of Sciences,
Ferdowsi University of Mashhad, Mashhad 91775-1436, Iran}

\begin{abstract}
We study Active Galactic Nucleus (AGN) activity in the fossil
galaxy cluster, RX J1416.4+2315.
Radio observations were carried out using Giant Metrewave Radio
Telescope (GMRT)
at two frequencies, 1420 MHz and 610 MHz. 
A weak radio lobe that extends from the central nucleus is detected in 610 MHz map.
  Assuming the radio lobe originated from the
central AGN,
we show the energy injection into the Inter Galactic Medium
(IGM) is only sufficient to heat up the central 50 kpc within
the cluster core,
while the cooling radius is larger ( $\sim$ 130 kpc). 
In the hardness ratio map, three low energy cavities have been identified. 
No radio emission is detected for these regions. We evaluated the power required to inflate the cavities and
showed that the total energy budget is sufficient to offset the radiative cooling. 
We showed that the initial conditions would change the results remarkably.    
Furthermore, efficiency of Bondi accretion to power the AGN has been estimated. 

\end{abstract}


\section{Introduction}
The Inter Galactic Medium (IGM) heating is one of the
challenging issues of
 observational cosmology. Many galaxy groups
and clusters have been found to contain hot intergalactic gas,
most of
which have cooling timescales longer than the Hubble time. The cooling time in the core of many galaxy groups and clusters
is however,
much shorter than the Hubble time. While it was argued that this gas
should
cool dramatically into very low temperatures \citep{Fabian84,
Sarazin86, Fabian91}, challenges were posed
by X-ray observations of the \xmm and the \chandra when they
found no
evidence of a catastrophic cooling suggesting that one or more
processes are stopping the gas from cooling below a certain
($\sim0.5
keV$) temperature. (see Peterson \& Fabian 2006 and references
therein).

Active Galactic Nucleus (AGN) is seen as a primary source
of heating mainly because it appears to be present in most of cool core 
galaxy groups and clusters \citep{Mittal09}. 
 AGNs interact with the intergalactic gas by sending radio jets, bubbles,
shocks and sound waves powered by supermassive black
holes (SMBHs) and heat the IGM via different mechanisms \citep{Gaspari11}. Amongst all the heating sources introduced up to now, bubble structures are the main supporting
evidence for the AGN heating \citep{Birzan04, Dunn04, Dunn05, 
Rafferty06, Nulsen07, Birzan08, Sanders09, Dunn10, O'Sullivan11,Cavagnolo10}.
Radio bubbles release huge amount of energy into the IGM in the form of mechanical work. The energy required to inflate a bubble is roughly $10^{55}$ erg in low mass systems up to $10^{61}$ erg in rich clusters.  
Recent studies of X-ray cavities show that mechanical heating provides sufficient energy to quench cooling rate in most of the cooling flow groups and clusters up to z$\sim$0.8 \citep{JHL12, JHL14}. 

The accretion flow onto the black hole prompt 
the relativistic jets and lobes into the IGM.
There are two modes of accretion into the SMBH that are assumed to contribute to the 
AGN power. The accretion of cold optically thick gas occurring at high accretion rate and hot
optically thin accretion of gas at the low accretion rate \citep{Yuan14}. The latter is believed to operate radio mode feedback in galaxy clusters and groups. Additionally,
galaxy simulations show that merger or instability caused by galaxy
interactions excite gas infall and fuel the black hole growth \citep{Springel05a,Springel05b}. Observational restrictions do not allow us to determine which type of accretion mode is induced by the galaxy mergers since the accretion mode  is mainly determined by the boundary condition of the gas at the Bondi radius \citep{Yuan14}.
When the black hole reaches a critical mass, the feedback blows out material from the central region and regulate the 
black hole growth. 
 A study by Ellison et al. 2011 shows a clear increase in the AGN fraction
in close pairs of galaxies. 
Therefore, merger of galaxies trigger AGN activity and consequently heat the IGM.
Moreover, the energy released by a galactic merger ($\sim 10^{64}$ erg) is enough to disturb cool cores
\citep{Markevitch99,
Gomez02, Markevitch07, McNamara07}. So the heating role of galaxy merger 
is important to account the heating and cooling balance inside the group.

Galaxy clusters and groups are evolving systems and galactic
merger
is a built-in process in these systems. However, there is a
class of galaxy groups
in which major mergers are thought to have taken place in time
scales much
larger than the return period of the AGN activities. Such galaxy groups are
known as fossil groups which are dominated optically by a single luminous elliptical
galaxy at the centre of an extended luminous X-ray emission
similar to
those seen in bright X-ray groups. The X-ray emission in fossils
is
usually regular and symmetric, indicating the absence of recent
group
scale merging. Apart from the observational indications pointing at the
early formation of fossil galaxy groups
\citep{kmpj06, kpj06, kpj07}, cosmological simulations
also show that fossil groups halos are formed relatively earlier \citep{dariush07, dariush10} than
the general population of the groups and clusters with a similar halo mass.

Two main advantages of studying AGN feedback in fossil groups are the following.
a) The central giant elliptical galaxy have not experienced a
recent galactic merger and, given
the absence of any bright nearby galaxies, is not subject to any major instability due to
interactions
with massive surrounding galaxies.
b) The cluster itself has not experienced any large scale merger which can disturb the formation of
the cool core.
Fossils are thus thought to be relatively simple laboratories to study AGN feedback.
This study is focused on the radio properties of one of the
most massive
fossil groups known in the local Universe, RX J1416.4+2315
(hereafter J1416) at $z\approx 0.137$.
This group contains one of the most massive giant ellipticals
known and has been studied as part of a volume
limited sample of spatially extended
X-ray sources compiled during the WARPS project (Wide Angle
ROSAT
Pointed Survey; Scharf et al. 1997; Jones et al. 1998; Perlman et al. 2002). It is
the
most X-ray luminous source in the sample of 5 fossil groups
studied by Jones et al. 2003, with a ROSAT estimated X-ray bolometric
luminosity of
$1.1 \times 10^{44}$ erg s$^{-1}$. In a deep optical image it
would be
classified as a galaxy group or poor cluster centred on an
extremely
dominant, luminous giant elliptical galaxy  \citep{kmpj06}.

In Section 2 we describe the radio observation and
Morphology. X-ray Morphology is presented in Section 3. Estimation of the heat
budget is described in
Section 4. The Synchrotron aging and Black hole accretion are presented in Section 5 and 6 respectively. Summary and discussion
 are given in Section 7. Throughout this paper a
$\Lambda CDM$ cosmology with
$\Omega_m=0.27$, $\Omega_\Lambda=0.73$ and $H_0=100 h$ km
s$^{-1}$ Mpc$^{-1}$ where
h=0.71 is assumed.\\

\section{Radio Observation and Morphology}
\noindent The cluster was imaged in two frequencies, 1420 MHz
and 610 MHz, with the
Giant Metrewave Radio Telescope (GMRT) to study a possible
correlation of the
spatial features of the IGM in radio and X-ray. The system
temperature, gain (K/Jy) and
full width at half maximum of the primary beam of the instrument in
1420 MHz are $\sim$ 76K, 0.22 and 24 \arcmin respectively. The same for 610
MHz are 92K, 0.32 and 54 \arcmin respectively. The source was
observed
on 4th and 5th July, 2005 for 10 hours in 1420 MHz and on 11th
and
12th July for 10 hours in 610 MHz.

\vspace*{0.2cm}

\noindent The data were reduced using
AIPS
(Astronomical Image Processing System). Bad data e.g.\ dead
antennas,
antennas with significantly lower gain, bad time, were
removed and the data was calibrated for flux and phase using
3C286 as
a primary calibrator in both frequencies and 1407+284 and
3C287
as secondary calibrators in 1420 MHz and 610 MHz
respectively.
Imaging was performed using the AIPS task 'IMAGR'. Polyhedron
imaging was
performed to take care of the well known problem of wide field
imaging in
low frequencies. Map rms $\sim$130 $\mu$J and $\sim$180 $\mu$J
were reached in
1420 MHz and in 610 MHz respectively.

\vspace*{0.2cm}

\noindent In 1420 MHz, a source was detected at the position of
RA: 14$^{h}$ 16$^{m}$ 27.47$^{s}$ and Dec: +23$^{0}$ 15$^{'}$
22.40$^{"}$. Peak flux of this source is 2.60
$\pm$ 0.13 mJy/beam. This is in agreement with the source from the
FIRST
catalog detected at the position of RA:14$^{h}$ 16$^{m}$
27.39$^{s}$ and Dec: +23$^{0}$ 15$^{'}$ 23.00$^{"}$,
with a peak flux of 2.36 $\pm$ 0.145 mJy/beam and with the source from the NVSS
catalog
at the position of RA:14$^{h}$ 16$^{m}$ 29$^{s}$ and Dec: +23$^{0}$15$^{'}$ 28$^{"}$, with a peak
flux of 3.1 $\pm$ 0.5 mJy/beam. High and low resolution images
of this source at 1420 MHz have been presented in figures 1 and 3. Details of
the maps are given in table 1.
\vspace*{0.2cm}

In 610 MHz, two sources were
detected in the high resolution image; a peak at the position of
RA: 14 16 27.35 Dec: +23 15 22.00 (hereafter central source) and a second peak 
at RA: 14 16 27.64 Dec: +23 15 18.00 (hereafter the radio
counterpart).
Peak flux for the central source is 4.29 $\pm$
0.13 mJy/beam
and for the radio counterpart is 2.00 $\pm$ 0.2 mJy/beam. High and low resolution
images of this source at 610 MHz are shown in figures 2 and 4. The nature of the radio
counterpart is unclear. There is no point source in the X-ray map
associated with this structure.
Given the alignment (along minor axis of the central giant
elliptical),
it can be a remnant from the previous outburst.
No counterparts are apparent either on the SDSS image 
(magnitude limit $m_{r}$ $<$ 17.7 )
or on 2MASS image (point-source limiting magnitude Ks $<$ 14.3). 
Thus in this study, we assume that this source is a radio lobe inflated by the central AGN.
 Given the
1420 MHz flux, the spectral index ($S {\propto} {\nu}^{-
\alpha}$) calculated for the central source is 0.83.
If the same spectral index is assumed for the radio lobe,
the source is expected to have a 3 $\sigma$ detection in 1420
MHz image,
given that the map rms is $\sim$130 $\mu$ Jansky.

\vspace*{0.2cm}

\noindent 

 The presence of diffused extended emission was 
checked by subtracting the high resolution 
integrated flux density of the source from the low resolution 
image. 
At 1420 MHz, no residual flux was found, indicating that the
emission is
dominated by a central compact source.
Subtracting FIRST flux density from NVSS flux density also
confirms it (Table 2).
At 610 MHz, there is $\sim$ 3 mJy residual that indicates the
presence of diffused emission.

\section{X-ray Morphology}

X-ray data analysis of J1416 has been reported in  Khosroshahi et al. 2006.
We used the results of Chandra and XMM-Newton observations in this study.
Throughout the paper, 
X-ray luminosities have been estimated by integrating over the surface brightness profile.
 A two-dimensional $\beta$-model fit to the Chandra data has been used for this aim.
In order to obtain the electron density $(n_{e})$, the deprojected
luminosity of 3D shells with the fixed radius was combined with the model emissivity
$\epsilon$ at the shell temperature
(and in the same energy band, 0.3$-$2.5 keV)
and the shell volume V, via
\begin{eqnarray}
L=n_{e} n_{H} V {\epsilon}
\end{eqnarray}
For the hydrogen number density we assumed
$n_{e}=1.17 n_{H}$. 
The deprojected temperature profile is reported in Khosroshahi et al. 2006.
 We estimated the pressure using the equation,
$p=\frac {kT\rho}{\mu m_{p}}$
for the perfect gas, with k being the Boltzmann constant, ${\mu}$ the mean
molecular weight, and $m_{p}$ the atomic mass constant.
For the gas density, we assumed ${\rho}=1.17 n_{e} m_{p}$ according to Reiprich et al. 2001.

X-ray soft (0.5$-$2 keV) diffuse emission from XMM-Newton with 610 MHz radio contours is shown in Fig. 5.
The diffuse emission image was produced
by adaptively smoothing an exposure-corrected image of the
emission detected by the three XMM-Newton EPIC cameras.
 There is a SW tail in the same direction of the radio lobe.  
 Compared to the X-ray emission at the same radius, the tail is
${\sim}$ 40 percent brighter with 6 ${\sigma}$ significance level.

Soft diffuse Chandra image (0.3-2.0 keV) is shown in Fig. 6. The point sources are removed and the emission has been smoothed with a two-dimensional Gaussian function, $\sigma =5$ pixels ( ${\sim}$ 5${"}$ kernel radius).
The image shows a depression in surface brightness map in the direction of the radio lobe. To assess the significance of this cavity, we averaged count per pixel over the cavity surface (marked as A) and the regions of the same radius from the center (marked as B). The mean value for the cavity is 0.085$\pm$0.009 (counts per pixel) comparing to 0.112$\pm$0.022 (counts per pixel) for the region B.  The surface brightness depression is weaker than the noise level in X-ray map. Thus the reliablity of the depresstion can not be investigated with the current X-ray data. Fig.7 shows the surface brightness profile of the entire cluster (open circle) and through the direction of radio lobe (filled square). The bin size has been adjusted to the resolution of the X-ray image. Y axis represents the mean of surface brightness in each bin with the standard deviation of each bin for the error bar. A small depression around $\sim$ 30 kpc is again uncertain due to the huge error bars.

The hardness ratio map is shown in Fig. 8. The soft image contains photons with energies from 400 eV to 900 e
V. The hard image consists of photons with energy ranging from 900 eV to 2 keV. Before deriving the hardness ratio map, we replaced the point sources by their surrounding
counts and applied smoothing on the hard and soft band with the scale of 8 $\times$ 8 binned image using
the csmooth task in CIAO. The corresponding
backgrounds were subtracted from the soft and hard images. The bubble like low energy cavities are
 seen in the south (marked as D and E) and north west (marked as C). In order to explore the reliably of these cavities, we compared the hardness ratio inside these regions with their surroundings by averaging count per pixel in each cavity and an annulus around it (not shown in Fig. 8 to prevent confusion). The level of deficiency of the cavities are presented in table 3 and show that surface brightness depressions are real in all cases.  Figs 9-11 show hardness ratio profile of the entire cluster (open circle) and through the direction of the cavities C, D and E, showing clear depressions around $\sim$ 20 kpc, $\sim$ 50 kpc and $\sim$ 80 kpc respectively. The bin size and the error bars are the same as Fig. 7. The cavities are possibly due to multiple
cycle of AGN outflows. However, no radio emission is detected in any of these
regions.

\section{AGN Heating}
X-ray observations of clusters and groups of galaxies reveal that most of the
cooling flow clusters and groups have cavities defined by
depression in their X-ray surface brightness. 
These cavities are filled with radio bubbles,
consist of relativistic particles and magnetic fields, when they observed at low frequencies. Such
observations are consistent with the current scenario in which AGN activity at
the centre of clusters and groups inflates the bubbles into the IGM, fed by
powerful radio jets. A large number of studies suggest that AGN as a prominent source of heating can quench
cooling rates through various processes \citep{Fabian06, Heinz06, McNamara07,
Blanton10}.
P dV work by the inflated bubbles on the surrendering can be a lower limit in the heat supply
\citep{Churazov02, Nusser06, Begelman01}. Other mechanisms will
boost the heating
and possibly explain the lack of cool gas in the clusters core.

From the cavity enthalpy we have
\begin{eqnarray}
H=E+PV={\frac{\gamma}{{\gamma}-1}}PV
\end{eqnarray}
In which we assumed the perfect gas relation for the internal energy of the cavity:
\begin{eqnarray}
E={\frac{1}{{\gamma}-1}}PV
\end{eqnarray}

Here, ${\gamma}=4/3$ is the adiabatic index for the relativistic
particles. In this study, we assume that the radio
bubble is formed near the centre of the group with over pressure
factor of ${p_0}/{p}$ (where ${p_0}$ is the initial pressure of the
bubble and $p$ is the
ambient pressure and also the bubble final
pressure, assuming the bubble pressure reaches the ambient
pressure in the last stage of expansion) and then moves to the observed position with the pressure and the volume
P and V. The enthalpy
change of the bubble during this process is given by

\begin{eqnarray}
{\Delta}H={\frac{\gamma}{{\gamma}-1}}PV[{({\frac{p_{0}}{p}}})^{1-1/{\gamma}}-1]
\end{eqnarray}

This is based on adiabatic assumption PV$^{\gamma}$=const and is
consistent with rapid expansion of the bubble.

Observation of J1416 in 610 MHz detected a radio source at the south-east of the cluster center but no such emission was detected at this position in 1420 MHz. 
The distance between the position of the peak flux density of the central source and the radio lobe is approximately 28 kpc. 
Absence of detection any radio lobe in 1420 MHz shows that this source has steep spectrum  up to
$\alpha \sim 2$, comparing to the central radio emission ($\alpha = 0.83$).
Steepening of about 0.5 is expected at high frequency as a result of synchrotron
ageing \citep{Feretti08}. Further discussion is presented in Section 5. 

The spatial resolution of the XMM-Newton is barely sufficient
for detecting any X-ray cavity at this distance from the central X-ray point source. Chandra
X-ray telescope which has a
much better spatial resolution shows a weak deficiency in the X-ray
surface brightness associated with the radio lobe (Figs 6 and 7).  The binning and smoothing
is minimal to enable us to see small scale surface brightness
variations in this map.  As we discussed  in Section 3, the depression is not reliable due to the huge error bars. Thus, we used radio data to evaluate bubble dimensions.

Here, we estimate the energy required to inflate the radio lobe (bubble heating) using the equation 4 and 
discuss the heating and cooling balance at the center of this cluster.
 The bubble has approximately an
elliptical shape with semi-major axis of 16 kpc and semi-minor
axis of 9 kpc (shown in Fig. 5). The extent of the source is calculated by fitting a Gaussian function to the image using the AIPS task JMFIT. We assume a mean radius of r=$\sqrt{ab}$ for the
bubble where $a$ and $b$ are
semi-minor and semi-major axis of the ellipse respectively. The ambient pressure is
$p=1.6-1.7$$\times$${10}^{-11}$dyn cm$^{-2}$ at the 28 kpc radius from the center (the current position of the radio bubble). The enthalpy change is ${\Delta}$$H=2.6 {\times}{10}^{57}erg$ assuming the overpressure
factor of 2 and ${\Delta}$$H=10.6 {\times}{10}^{57}erg $ for the overpressure factor of 10. Dividing the
enthalpy changes by the time of releasing this energy we reach the bubble luminosity.
Using three
estimates for the time introduced by B\^\i rzan et al. 2004, we evaluated
heating luminosities listed in table 4. The sound crossing time, $t_{c}= \frac{R}{\sqrt{\frac{\gamma kT}{\mu m_{H}}}}$ is
the time for the bubble to move to the current place with the sound
speed in the group core. The buoyant rise time, $t_{b}= \frac{R}{\sqrt{\frac{2gV}{SC}}}$ is the time which bubble moves out
buoyantly and the refill time, $t_{r}= 2 \sqrt{\frac{r}{g}}$ that is the time taken to
refill the volume
displaced by the bubble during the movement. 
Here, R is the distance between galaxy center and radio lobe. V and S
are the volume and cross section of the bubble respectively. C is the drag coefficient. We assumed C=0.75 (Churazov et al. 2001).
g is  gravitational acceleration that is approximated by the relation $g \approx 2 {\sigma}^{2}/R$ where $\sigma$ is the velocity dispersion of the galaxy group ($\approx$ 694 km s$^{-1}$). r is the mean radius of the bubble which we discussed earlier. 

The X-ray study of J1416 \citep{kmpj06} shows that the gas
within 130 kpc have cooling time shorter than the Hubble time
while there is only a small temperature drop in the core of
galaxy group (r ${\textless}$ 50 kpc) where the cooling time is ${\sim}$ 5
Gyr. The energy loss
by X-ray emission is $L_x=2.8{\times}{10}^{42}$ erg
s$^{-1}$ within 50 kpc and $L_x=2.5{\times}{10}^{43}$ erg
s$^{-1}$ within 130 kpc. The mechanical luminosity of the radio lobe is
${L_{mech}}=2.2{\times}{10}^{42}erg s^{-1}$, assuming overpressure factor of 2 and 
${L_{mech}}=1.2{\times}{10}^{43}erg s^{-1}$ for the overpressure
factors of 10 ( Table 4). Thus, the estimated
mechanical heating appears to be only sufficient to heat up the cool
core of the group and at higher radius, other sources of heating are also required. 
The fact that AGN feedback can only touch the inner radius preserves 
self-similarity in galaxy groups and clusters. A recent study by Gaspari et al. 2014 
shows that breaking the self-similar scaling relations
destroys cool core in galaxy groups and clusters and changes all systems into non-cool-core objects
that is not consistent with X-ray observations.

We note that the highest angular resolution achieved in 610 MHz map
is approximately equal to the size of the radio bubble.
Since we use the radio bubble size to calculate the volume (equation 3), 
the estimated mechanical work will be an upper limit for that.
For the B\^\i rzan et al. 2004 sample, the cavities dimensions vary
from tens of kpc in massive clusters to few kpc in low mass
systems.
Thus, our estimate for the bubble size ($\sim $12 kpc) is
consistent with a typical expected size.

 The properties of the core can be seen better in the hardness ratio map (Fig. 7) which shows asymmetric distribution of soft and hard regions. The bubble like low energy cavities are marked in the map. There is no detected radio counterpart for these structures possibly due to the synchrotron aging. Future low frequency ($<$ 610 MHz) radio observation of this object with the high sensitivity  will discover the nature of these cavities. All the cavities are within 130 kpc (cooling radius) from the cluster center. Assuming they emerge from multiple cycle of AGN outburst, we estimated the mechanical power for each cavity listed in table 3. The sound crossing time, the cluster pressure at the position of the cavity and over pressure factor of 10 (the maximum mechanical work) have been used to estimate the power. The total energy content of the cavities (A, C, D and E) is $\Sigma PV=9.6 \times 10^{58}$ erg corresponds to the total mechanical luminosity of ${L_{mech}}=1.3{\times}{10}^{44}erg s^{-1}$.
Therefore, the heating power is completely sufficient to quench cooling rate within 130 kpc in the cluster core. By changing the initial conditions of inflated bubble and applying overpressure factor of 2 (the minimum mechanical work), the heating power is hardly sufficient to stop the gas from cooling, ${L_{mech}}=1.9{\times}{10}^{43}erg s^{-1}$. We will discuss it in section 7.

In addition to the mechanical heating of the radio lobe, dissipation of shock waves and sound waves
produced by AGN outbursts is another source of heating \citep{McNamara07}. This is
illustrated by cavities, weak shocks and filamentary structures
produced by series
of AGN outbursts in X-ray image of M87 Galaxy \citep{Forman07}.
For J1416, both Chandra and XMM-Newton in soft band (0.5-2.0 kev) show relatively smooth emission
except at the core where 
an X-ray extension appears in the south-east and the same direction of the radio
bubble in the XMM-Newton map (Fig. 5). It is probable that this
structure is related to the AGN
outburst or the result of low resolution and
high smoothing of
XMM-Newton map. Any firm conclusion about the existence of weak shock will require a higher quality of X-ray data.

Observations of a large sample of clusters and groups have
helped to deduce a correlation between the radio luminosity and the jet
power. B\^\i rzan et al. 2004 studied a sample of galaxy clusters
and groups with clear X-ray cavities and derived a relation between cavity power
and total radio luminosity,
showing that the ${L_{mech}}$ is approximately proportional to
the $\sqrt{L_{radio}}$ but radio luminosity is on average,
smaller by a factor of 100.

We use radio flux density of the central source $S_{\nu_0}$ at
the frequency ${\nu_0}=1.4GHz $ to evaluate the total radio luminosity with the assumption of power low
spectrum between ${\nu_1}=10MHz$
and ${\nu_2}=5000MHz$ and the spectral index of ${\alpha=0.83}$.
Using the relation
\begin{eqnarray}
{L_{radio}}={4{\pi}}{D_{L}}^{2
}{S_{{{\nu}_0}}}{\int_{\nu_1}^{\nu2}}
({\frac{\nu}{{\nu}_0}})^{-{\alpha} } d{\nu}
\end{eqnarray}

we approximate the total radio luminosity to be
$1.1{\times}{10}^{40}$ erg s$^{-1}$. The corresponding 
mechanical luminosity according to  B\^\i rzan et al. 2004 relation is $4.1{\times}{10}^{42{\pm}5}$ erg
s$^{-1}$,  consistent with our results in Table 4.

\section{Synchrotron ages and the magnetic field}
Assuming that the radio lobe is in pressure balance with the
environment, the gas pressure estimated from the X-ray
observation should be equal to the lobe internal pressure. 
The radio lobe pressure is the sum of particles and magnetic 
field pressure and is evaluated using the bolometric radio luminosity of the lobe.

\begin{displaymath}
P_{th}=P_{p}+P_{B}
 \end{displaymath}
\begin{eqnarray}
={\frac{(1+k)}{3V{\phi}}}C_{12}({\alpha};{\nu_{1}};{\nu_{2}})B^{-3/2}L_{radio}+{\frac{B^{2}}{8{\pi}}}
\end{eqnarray}

Here, $P_{th}$ is the gas thermal pressure. $k$ is the
coefficient that relates the energy of positive particles to
the energy of electrons. ${\phi}$ is the filling factor of magnetic
field. $C_{12}$ is a function of emitting first and last frequencies
and spectral index, introduced by Pacholczyk 1970. We assume the radio lobe consists of
relativistic electrons and
positrons $(k=0)$, with a filling factor of ${\phi=1}$.
The applied minimum and maximum cut off for frequencies are
${\nu_1=10MHz}$ and ${\nu_2=100GHz}$.
Following this method, we found two solutions for the magnetic
field of the lobe, $B_1=20{\mu}G$ and $B_2=0.5{\mu}G$. $B_2$
seems to be too small according to the magnetic field strength
in radio lobes \citep{Stawartz06}. Thus, we adopted $B_1=20{\mu}G$
that shows magnetic
pressure is dominated inside the radio lobe.

Unlike the radio emission at 610 MHz, there is no radio
lobe in 1420 MHz map. Assuming the same spectral index for the central source and the radio lobe,
 the emission of the radio lobe at 1420 MHz would be larger than 3$\sigma$ of the map.
Thus, there are two
possibilities for
explaining the lack of radio lobe detection at high frequency. Firstly, difference in electron energy distribution of the lobe and the central source causes a
steeper spectral index for the radio lobe. secondly,
the steepening emerges from synchrotron ageing. The former is not probable, since both of the radio emissions have been originated from the central AGN while the latter is a common phonamonon observed in radio lobes. The energy of the synchrotron electrons in the
plasma decreases with time because of their emission.
This phenomenon defines a critical time ($t_{syn}$) in which
particles lose most of their energies, and produces a break at the
synchrotron spectrum for the frequencies beyond the critical
frequencies (${\nu}>{\nu}_c$). The break frequency (${\nu}_c$)
is related to the
synchrotron age via the relation
\begin{eqnarray}
{t_{syn}}=1060{\frac{B^{1/2}}{{B}^2+{2/3}{B_{CMB}}^2}}{[{\nu}_c(1+z)]}^{-1/2}Myr
\end{eqnarray}
Here, B is in ${\mu}G$. ${\nu_c}$ is in GHz and
${B_{CMB}}=3.25{(1+z)}^2{\mu}G$ is the equivalent Cosmic
Microwave Background magnetic field. The electron energy also
losses due to the inverse Compton scattering of CMB photons (see Feretti \& Giovannini 2008 for details).
The absence of any radio lobe at 1420 MHz
indicates a spectrum break below this frequency, resulting
$t_{syn}{\gtrsim} 1 {\times} {10}^{7} yr $. This is consistent
with the three estimated age for the bubble in Section
4.

\section{Accretion and blackhole growth}
In addition to the energetic considerations around heating and 
cooling balance at the core of galaxy groups and clusters,
the energetic balance of AGN feeding and feedback is 
required to complete the cycle of galaxy formation and
growth. The SMBH accretion by the hot or cold gas 
are widely discussed in the literature \citep{Pizzolato05, McNamara11, Gaspari13, Yuan14}.
The rotational energy of a rapidly spinning black hole can also power AGN activity \citep{McNamara09}. 
In this section, we approximate Bondi accretion \citep{Bondi52}, a spherical symmetric accretion of
hot gas into the central black hole, as a fueling way to know whether or not it is sufficient to
feed the central black hole. The correlation of Bondi accretion and jet power 
has been shown in the previous studies \citep{Rafferty06, Allen06}. 

The accretion rate is related to the temperature, density and
the central black hole mass via the relation:
\begin{eqnarray}
{\dot{M}_{bondi}}=1.66{\times}{10^{-7}}{M_{7}}^{2 }{T_{2}}^{-3/2
}{n_{0.1}}{M_{sun}}{yr}^{-1}
\end{eqnarray}
we evaluate black hole mass using the correlation between the black hole
mass and $K_{s}$ luminosity \citep{Marconi03}:
\begin{eqnarray}
{log{\frac{M_{BH}}{M_{sun}}}}=8.21+1.13[{log{\frac{L_{K}}{L_{sun}}}}-10.9]
\end{eqnarray}

The estimated black hole mass is $M_{BH}=2.4 \times
10^{9}M_{Sun}$. Using the core
density ${n}_{e}=0.002$ cm$^{-3}$ and temperature $T=4$ keV, we
found the accretion rate
of $\dot{M}=6.7{\times}10^{-5}$M$_{sun} yr^{-1}$ , corresponding to
the Bondi accretion of
$P_{bondi}={\epsilon}{\dot{M}}{c}^{2}=3.8{\times}{10}^{41}$erg s$^{-1}$. In this estimation, we
assumed an efficiency factor of ${\epsilon}=0.1$ for the
accretion. The calculated Bondi
accretion power is an order of magnitude lower than the jet
power (section 4). Note that 
the accretion happens within Bondi radius. For a black hole mass $M_{BH}$ 
and accreting gas temperature T, Bondi radius $R_{bondi}$ is:
\begin{eqnarray}
R_{bondi}=0.031 ({\frac{kT}{keV}})^{-1} ({\frac{M_{BH}}{10^{9}M_{sun}}}) kpc
\end{eqnarray}
In J1416 galaxy cluster, this radius is $\sim$ 0.02 kpc. 
Spatial resolution of X-ray data do not allow us to resolve this region. Therefore, 
the true Bondi power must be higher than this value.
Rafferty et al. 2006 discussed this correction and showed that having a bigger gap between Bondi radius and 
the radius which Bondi power have been estimated within, we observe higher cavity-Bondi power ratio. 
In order to correct this effect, they assumed a power law density profile and predict the Bondi accretion rate.
Applying this correction for J1416 shows that cavity power is consistent with Bondi power.

\section{Summary and discussion}
We carried out radio observations of RX J1416.4+2315, a
massive fossil galaxy group, to study AGN activity and IGM heating in this object.
Fossil groups are thought to be ideal environment to study IGM
heating since they show no sign of group scale mergers or major
mergers of the dominant galaxy. The radio observations were
performed using GMRT radio array at 610 MHz and 1420 MHz.

The radio emission of the central BCG  is detected at both
frequencies, showing spectral index of 0.83. In 610 MHz, a radio lobe is detected
 in the SE of the central source. We evaluate mechanical work that is done by the radio bubble on the
IGM. If the bubble blows with small overpressure (${\frac{p_{0}}{p}}$=2), the heating
power will be $\sim 3{\times}{10}^{42}$erg s$^{-1}$ which is
sufficient to quench the cooling within $\sim$50 kpc radius.
 The temperature profile of J1416 shows a
slight drop in this region.
Assuming that the bubble initial pressure
is so high (${\frac{p_{0}}{p}}$=10),
the heat supply will be $\sim 1{\times}{10}^{43}$erg s$^{-1}$.
The total heating power that is required to completely stop
cooling within the cooling radius ($\sim$130 kpc) is
$2-3{\times}{10}^{43}$erg s$^{-1}$. Thus, the PV work  
is not sufficient for the entire cooling flow region. 
 In addition to that, x-ray hardness ratio map show three clear cavities within the cooling radius. 
They might be the results of previous AGN outbursts. By taking into account all sources, the heating power will be $\sim {10}^{44}erg s^{-1}$, that is quite sufficient to offset cooling flow. 

We estimate Bondi accretion rate of $ 3.8{\times}{10}^{41}$erg
s$^{-1}$ for this system. The accretion power is significantly smaller than
the jet power but such a difference emerges from the poor X-ray resolution. 
Applying the correction for the Bondi radius, we showed that Bondi accretion and jet power are consistent.

We note that there are uncertainties in estimating the
mechanical heating for the various reasons.

- The energetic content of the radio lobe is a function of
adiabatic index (Eq. 3), depends on whether the bubble
is filled with relativistic or non-relativistic plasma.

- Large uncertainty in evaluating the heating power is clear due to the method we used.
Total enthalpy and enthalpy changes are used for this aim.
Nusser et al. 2006 calculated the heating rate by introducing three phases of
injected bubble, buoyant bubble, and destroyed bubble. They derived precise relations while
applying the observable parameters into the deduced relations is not possible because of the low resolution of the X-ray data.
Moreover, assumptions for the initial conditions e.g bubble overpressure factor
 change the results by an order of magnitude. 

- Uncertainty in the definition of cooling radius based on the
cooling time,
temperature drop or significant classical mass deposition rate
for estimating
the X-ray loss results huge difference in the estimated cooling power.

\clearpage

\begin{figure*}
\centering
 \includegraphics[ scale=0.4]{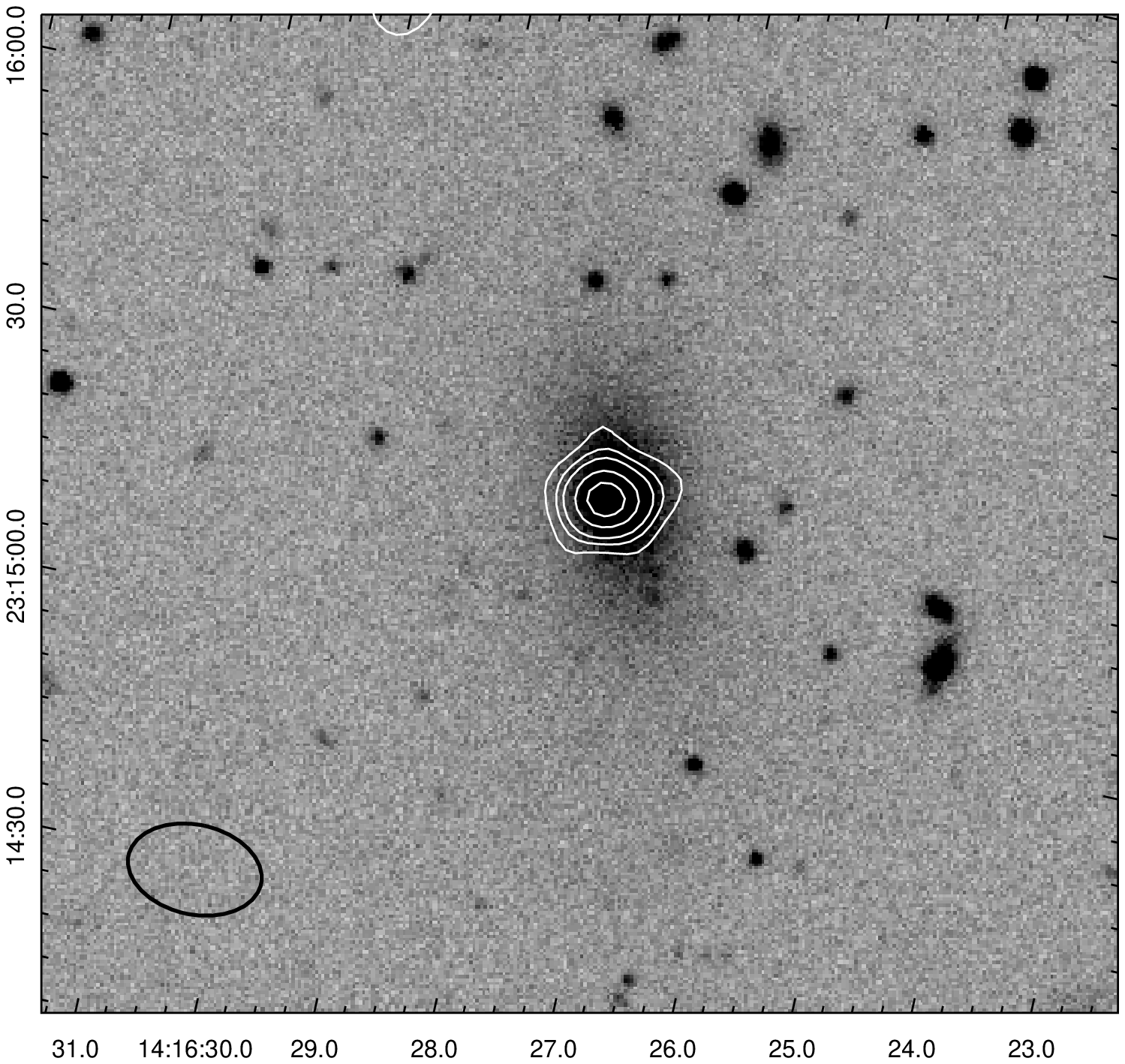}
\figcaption{The 1.4 GHz high resolution radio map of the source
overlaid on the SDSS optical image of the central galaxy, RX
J1416.4+2315.
Contour levels of 
 0.45 0.75 1.05  1.65  2.25 mJy  are shown. The synthesized beam is shown on the bottom-left of the image.
\label{Highfreqradio}}
 \end{figure*}
\clearpage

\begin{figure*}
\centering
 \includegraphics[ scale=0.4 ]{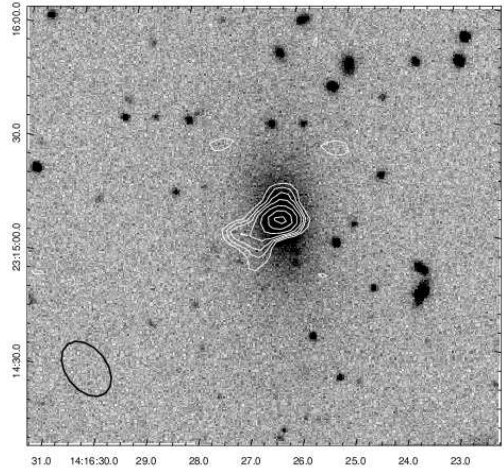}
\figcaption{The 610 MHz high resolution radio map of the source
overlaid on the SDSS optical image of the central galaxy, RX
J1416.4+2315.
Contour levels of 
 0.66, 0.88, 1.42, 2.0, 3.0, 4.0 mJy  are shown. The synthesized beam is shown on the bottom-left of the image.
\label{Lowfreqradio}}
 \end{figure*}
\clearpage

\begin{figure*}
\centering
 \includegraphics[scale=0.4]{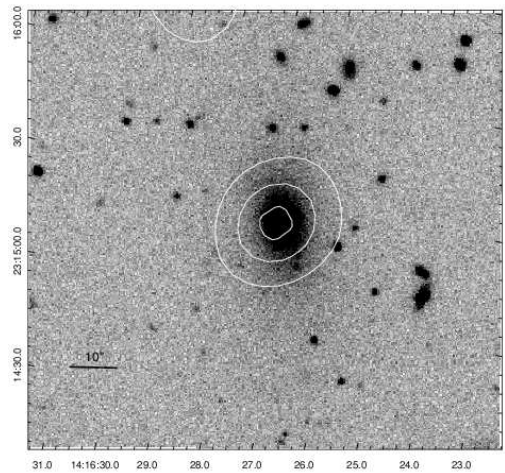}
\figcaption{The 1.4 GHz low resolution radio map of the source
overlaid on the SDSS optical image of the central galaxy, RX
J1416.4+2315.
Contour levels of 
 0.0015 0.0025 0.0032 Jy  are shown.
\label{Highfreqradio}}
 \end{figure*}
\clearpage

\begin{figure*}
\centering
 \includegraphics[scale=0.4]{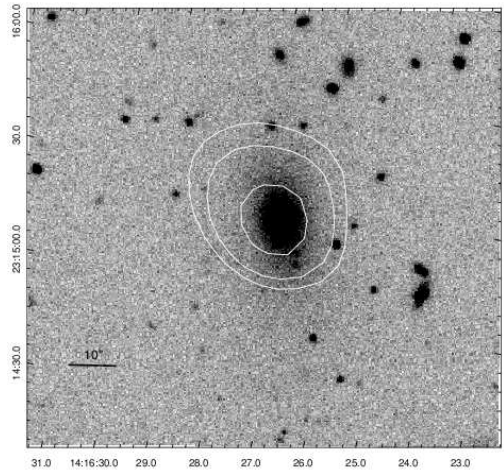}
\figcaption{The 610 MHz low resolution radio map of the source
overlaid on the SDSS optical image of the central galaxy, RX
J1416.4+2315.
Contour levels of 
 0.004 0.005 0.007, 0.0079 Jy  are shown.
\label{Highfreqradio}}
 \end{figure*}

\clearpage

\begin{figure*}
\centering
 \includegraphics[scale=0.8]{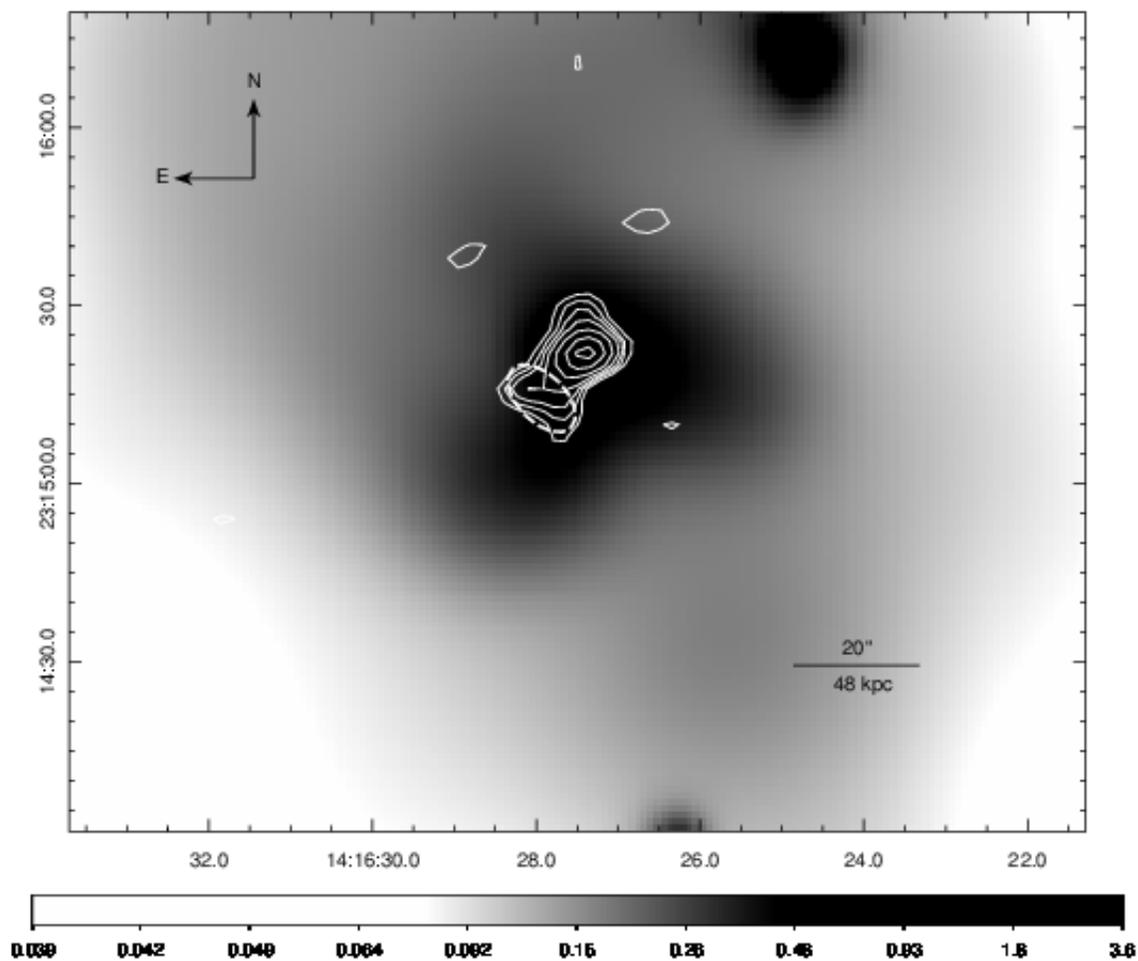}
\figcaption{The X-ray smooth emission ($\sim$10'' kernel radius) from the XMM-Newton
observations of cluster RX J1416.4+2315 with 610 MHz radio contours. The dashed ellipse demonstrates the bubble size used for
the volume estimation. 
\label{XMMradio}}
 \end{figure*}

\clearpage

\begin{figure*}
\centering
 \includegraphics[scale=0.8]{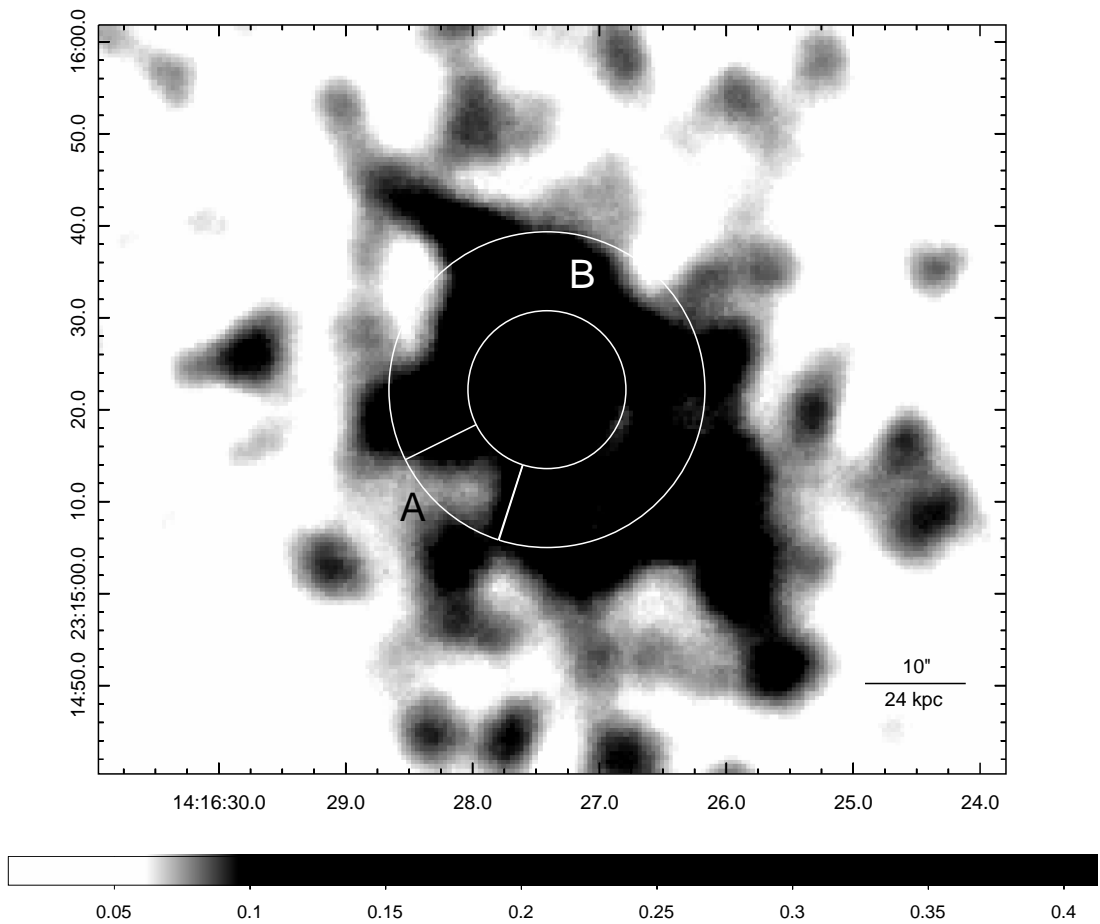}
\figcaption{The X-ray emission of the cluster from Chandra observations, There is a depression in region A where the radio contours show the existence of a radio lobe. The image has been smoothed with 5'' kernel radius.
\label{XMMradio}}
 \end{figure*}

\clearpage

\begin{figure*}
\centering
 \includegraphics[scale=0.8]{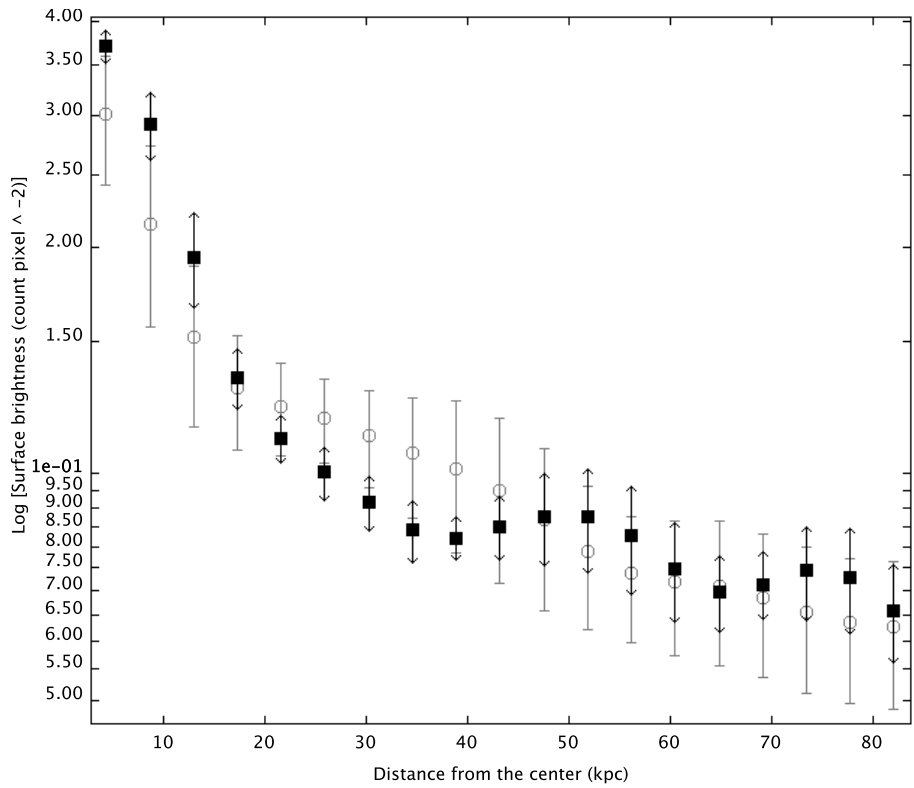}
\figcaption{The X-ray surface brightenss profile of the cluster from Chandra observations (open circle) and the same trough the direction of the radio lobe (filled square), There is a weak depression around $\sim$ 30 kpc which is uncertain according to the error bars. 
\label{XMMradio}}
 \end{figure*}

\clearpage

\begin{figure*}
\centering
 \includegraphics[scale=0.8]{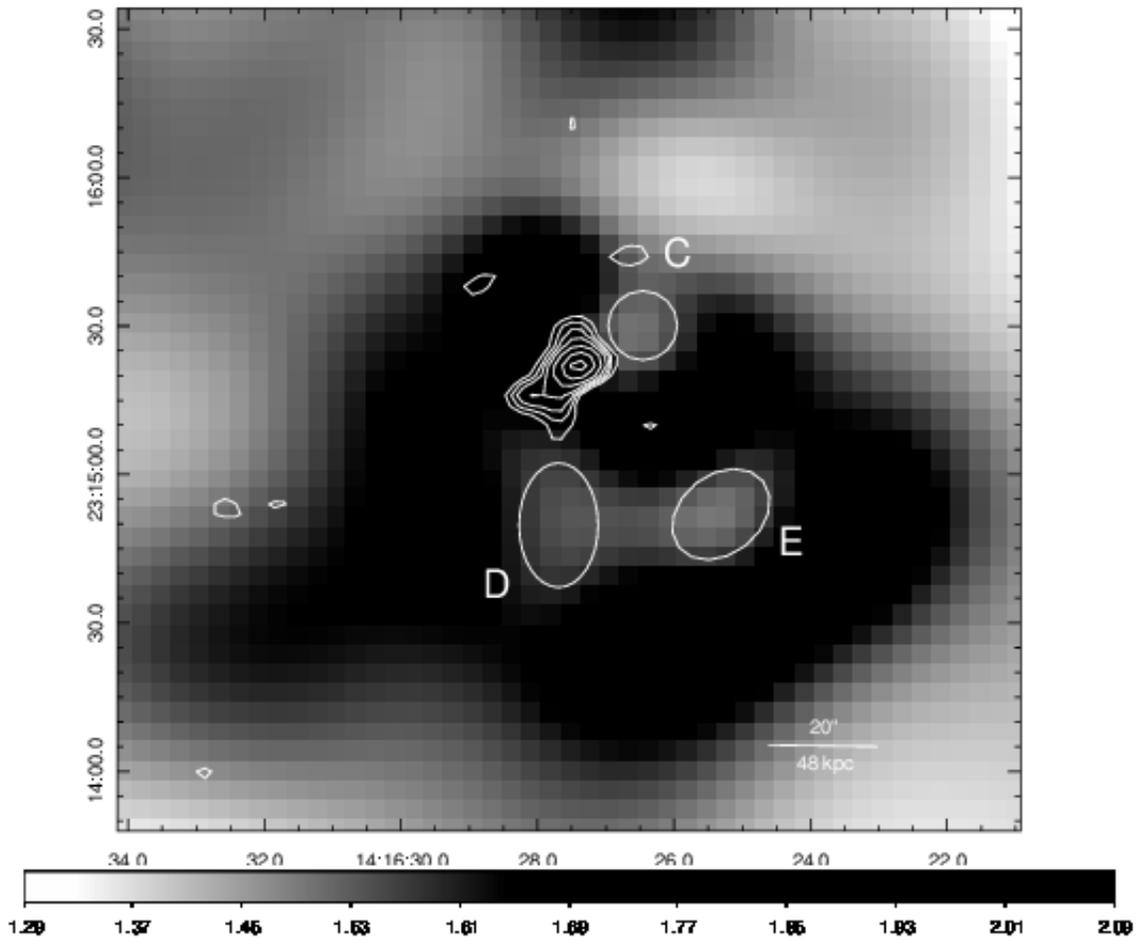}
\figcaption{The hardness ratio map from the XMM-Newton observations, Three cavities are marked in this map (C, D and E). 610 MHz radio contours are also shown. The image has been smoothed with 10'' kernel radius.   
\label{XMMradio}}
 \end{figure*}

\clearpage

\begin{figure*}
\centering
 \includegraphics[scale=0.8]{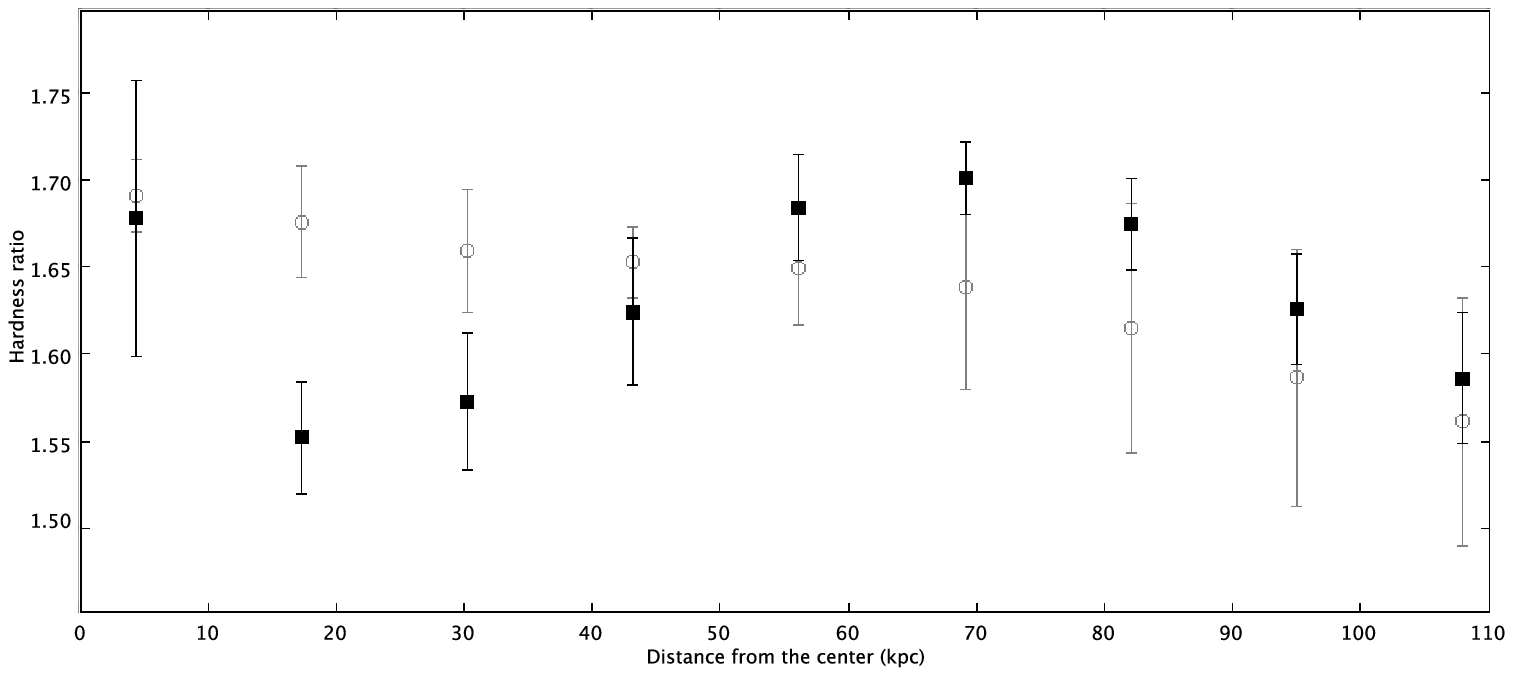}
\figcaption{The hardness ratio profile of the cluster (open circles) and the same trough the direction of the cavity C (filled square). There is a real depression around $\sim$ 20 kpc consistent with the cavity position.   
\label{XMMradio}}
 \end{figure*}

\clearpage\begin{figure*}
\centering
 \includegraphics[scale=0.8]{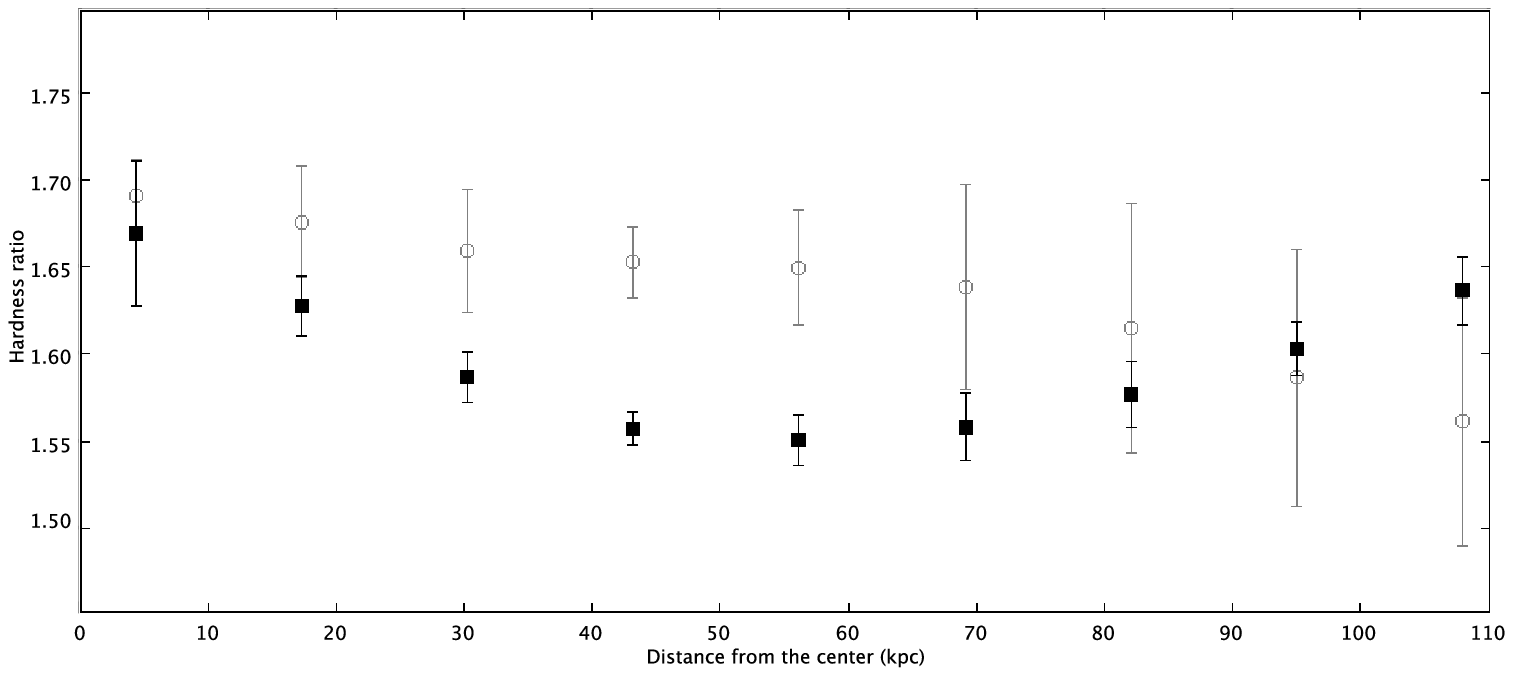}
\figcaption{The hardness ratio profile of the cluster (open circles) and the same trough the direction of the cavity D (filled square). There is a real depression around $\sim$ 50 kpc consistent with the cavity position.   
\label{XMMradio}}
 \end{figure*}

\clearpage\begin{figure*}
\centering
 \includegraphics[scale=0.8]{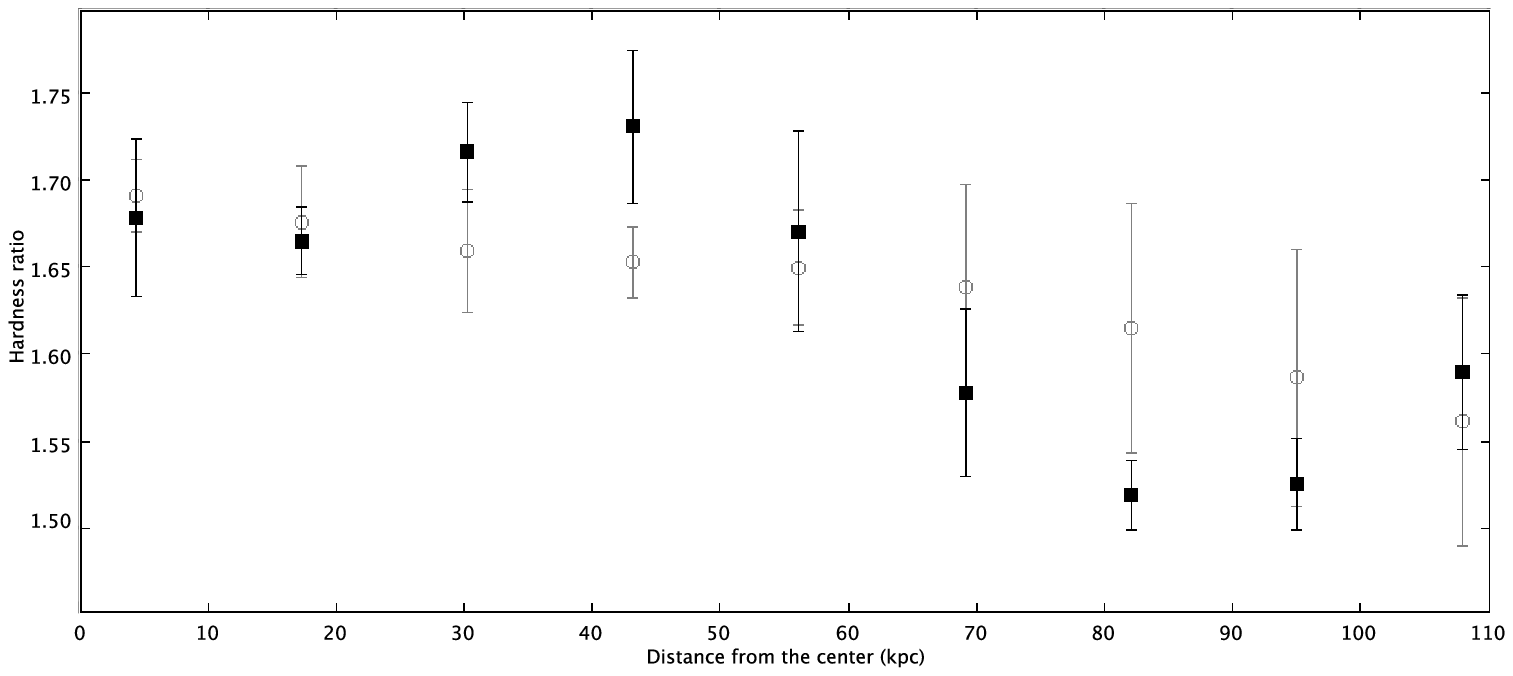}
\figcaption{The hardness ratio profile of the cluster (open circles) and the same trough the direction of the cavity E (filled square). There is a real depression around $\sim$ 80 kpc consistent with the cavity position.   
\label{XMMradio}}
 \end{figure*}

\clearpage

\begin{deluxetable}{crrrrr}
\footnotesize
\tablecaption{Details of the radio measurements of
RXJ1416.4+2315.
\label{table1}}
\tablewidth{0pt}
\tablehead{
\colhead{Frequency} & \colhead{Position} & \colhead{Peak Flux} &\colhead{Integrated Flux} &
\colhead{Map rms}  & \colhead{Resolution}    \\
\colhead{(MHz)} & \colhead{J2000} & \colhead{(mJy)} &
\colhead{(mJy)} &
\colhead{(mJy)}  & \colhead{(arcsec)}  
} 

\startdata
610 & 14$^{h}$ 16$^{m}$ 27.357$^{s}$ &4.29 $\pm$ 0.13&7.10 $\pm$0.33& 0.18 & 8.03 $\times$ 5.36\nl
      &   +23$^{0}$ 15$^{'}$ 22.00$^{"}$  &  & &&\nl
610 &14$^{h}$ 16$^{m}$ 27.647$^{s}$ &2.00 $\pm$ 0.20 &2.08 $\pm$0.20&0.18 & 8.03 $\times$ 5.36\nl
   &+23$^{0}$ 15$^{'}$ 18.00$^{"}$&&&&\nl
610 & 14$^{h}$ 16$^{m}$ 27.502$^{s}$ &7.60 $\pm$ 0.65&12.41
$\pm$ 1.60& 0.98 & 40.39 $\times$ 34.28\nl
      &   +23$^{0}$ 15$^{'}$ 24.00$^{"}$ &  & &&\nl
1420 & 14$^{h}$ 16$^{m}$ 27.473$^{s}$ &2.60 $\pm$ 0.13&3.55
$\pm$ 0.27&0.13 & 8.30$\times$6.96 \nl
&+23$^{0}$ 15$^{'}$ 22.40$^{"}$&&&&\nl
1420 &14$^{h}$ 16$^{m}$ 27.502$^{s}$ &3.40 $\pm$ 0.43&3.48 $\pm$0.76&0.36 & 32.98$\times$29.78 \nl
&+23$^{0}$ 15$^{'}$ 22.00$^{"}$&&&&\nl
\enddata


\end{deluxetable}
\clearpage
\begin{deluxetable}{crrr}
\footnotesize
\tablecaption{1420 MHz radio measurements of RXJ1416.4+2315.
\label{table2}}
\tablewidth{0pt}
\tablehead{
\colhead{Catalog} & \colhead{Peak Flux} & \colhead{Integrated
Flux} & \colhead{Map rms} \\
\colhead{} & \colhead{(mJy)} & \colhead{(mJy)} & \colhead{(mJy)}} 

\startdata
GMRT & 2.60 &3.55&0.13 \nl
FIRST & 2.36  & 3.39 & 0.145\nl
NVSS &3.1  &3.6 &0.5\nl
\enddata


\end{deluxetable}
\clearpage

\begin{deluxetable}{crrrr}
\footnotesize
\tablecaption{Hardness ratio map, cavity characteristics.
\label{table3}}
\tablewidth{0pt}
\tablehead{
\colhead{Cavity} & \colhead{Cavity hardness ratio} & \colhead{Anullus hardness ratio} & \colhead{$t_{c}$} & \colhead{$P_{c}$}  \\
\colhead{} & \colhead{} & \colhead{} & \colhead{${10}^{7}yr$} & \colhead{${10}^{43}erg s^{-1}$} } 

\startdata
C & 1.519 $\pm$ 0.015 &1.607 $\pm$ 0.060 & 3.9 & 4.3\nl
D & 1.551 $\pm$ 0.015 &1.662 $\pm$ 0.073 & 8.6 & 4.3\nl
E & 1.548 $\pm$ 0.036 & 1.698 $\pm$ 0.053 & 11.1 & 3.6\nl
\enddata
\end{deluxetable}
\clearpage

\begin{deluxetable}{crrrrrrrrr}
\tabletypesize{\scriptsize}
\tablecaption{energy injection in the core of RXJ1416.4+2315
\label{table2}}
\tablewidth{0pt}
\tablehead{
\colhead{${p_0}/{p}$\tablenotemark{a}} & \colhead{$t_{r}$} &
\colhead{$t_{b}$} & \colhead{$t_{c}$} &
\colhead{${\Delta}H$} & \colhead{$P_{r}$} & \colhead{$P_{b}$}
&\colhead{$P_{c}$} & \colhead{$L_{(x<50kpc)}$} &
\colhead{$L_{(x<130kpc)}$}   \\
\colhead{} & \colhead{${10}^{7}yr$} & \colhead{${10}^{7}yr$} &
\colhead{${10}^{7}yr$} &
\colhead{${10}^{57}erg$} & \colhead{${10}^{42}erg s^{-1}$} &
\colhead{${10}^{42}erg s^{-1}$} &
\colhead{${10}^{42}erg s^{-1}$} & \colhead{${10}^{42}erg
s^{-1}$} &
\colhead{${10}^{43}erg s^{-1}$} 
} 

\startdata
2&3.7&3.0&2.7&2.6&2.3&2.7&3.1&2.8&2.5\nl
10&3.7&3.0&2.7&10.6&9.2&11.2&12.5&2.8&2.5\nl
\enddata

 
\tablenotetext{a}{Over pressure factor of injected lobe}

\end{deluxetable}

\end{document}